\documentclass[twocolumn,jcp]{revtex4}

\usepackage{graphicx}

\begin{document}

\title{First Principles NMR Signatures of Graphene Oxide }

\author{Ning Lu}
\author{Ying Huang}
\author{Haibei Li}
\author{Zhenyu Li}
\thanks{Corresponding author. E-mail: zyli@ustc.edu.cn}
\author{Jinlong Yang}

\affiliation{Hefei National Laboratory for Physical Sciences at
     Microscale,  University of Science and Technology of
     China, Hefei,  Anhui 230026, China}

\begin{abstract}
Nuclear magnetic resonance (NMR) has been widely used in the
graphene oxide (GO) structure study. However, the detailed
relationship between its spectroscopic features and the GO
structural configuration has remained elusive. Based on first
principles $^{13}$C chemical shift calculations using
the gauge including projector augmented waves (GIPAW) method, we provide a
spectrum-structure connection. Chemical shift of carbon is found
to be very sensitive to atomic environment, even with an identical
oxidation group. Factors determining the chemical shifts for epoxy
and hydroxy groups have been discussed. GO structures previously
reported in the literature have been checked from the NMR point of view.
The energetically favorable hydroxy chain structure is not
expected to be widely existed in real GO samples according to our NMR
simulations. The epoxy pair we proposed
previously is also supported by chemical shift calculations.
\end{abstract}

\maketitle

\section{introduction}
Graphene has been the rising-star material in the past several
years, due to its unique electron structure and its great application
potential. \cite{Novoselov200466, Geim200783, Katsnelson200720,
Castr200909} However, its large-scale production remains a big challenge.
Reduction of graphene
oxide (GO), which is obtained by oxidation and exfoliation of
graphite, \cite{Dreyer201040} provides a promising way to obtain
graphene or chemically modified graphene massively.
\cite{Stankovich200682, Gilje200794, Park200917, Gao200903} Solution
based chemical processes are expected to be applicable to many kinds
of applications, which makes GO itself an
important material. \cite{Dikin200757, Wu200801, Wang200995, Scheuermann200962,
Eda201005, Lightcap2010xx}

Despite a big effort from both experimentalists \cite{Mermoux199169,
He199654, Lerf199757, Lerf199877, Szabo200640, Cai200815, Gao200903}
and theoreticians, \cite{Boukhvalov200897, Lahaye200935, Li200920,
Yan200902, Zhang200905, Gao200963} the precise chemical structure of
GO is still not clear. The main reason for the difficulty to
understand its structure is the amorphous and nonstoichiometric
nature of GO. At the same time, different samples have different levels of oxidation, which makes things even more complicated.
Nevertheless, some fundamental structural features of GO, as
proposed by Lerf and coworkers, \cite{Lerf199877} have been widely
accepted. In this so-called Lerf mode, hydroxy (-OH) and epoxy (-O-) groups spread across the
graphene planes, while carboxylic acid groups (-COOH)
exist at edge sites, possibly in addition to the keto groups.

Aside from this general picture, there are still many open
questions about the GO structure. For example, how are the hydroxy
and epoxy groups distributed on the GO plane? Do they aggregate
together or avoid each other? Is there any other new species in GO,
especially for highly oxidized
samples? Are $sp^2$ carbon atoms clustered in aromatic forms?

It is very difficult to answer these questions based only on
experimental raw data. Therefore, theoretical modeling has played an
important role in GO structure study. Based on first principles
energetics, many GO structure models have been proposed. Adapting
a 2$\times$2 unit cell, Boukhvalov et al. \cite{Boukhvalov200897}
and Lahaye et al. \cite{Lahaye200935} systematically studied the
possible atomic configurations with different ratios of epoxides to
alcohols at different degrees of oxidation. Based on a study with
similar strategy, Yan et al. \cite{Yan200902} emphasized that it is
favorable in energy to form hydroxyl chains with hydrogen bonds. In a GO
hydrogen storage study, Wang et al. \cite{Wang200995} constructed a
GO model with a building block which has the lowest energy at a very
low oxidation group coverage.

Although it is very straightforward to conduct in theoretical
study, the power of energetics analysis is expected to be
limited by the complexity of the GO potential energy surface,
especially when artificial periodic boundary condition must be
adopted with a small unit cell. In contrast, computational spectroscopy provides
information which can be directly compared with experiments. Therefore, it
is a powerful alternative in nanostructure studies. For
example, using Raman spectrum simulation, Kudin et al.
\cite{Kudin200836} proposed an alternating single-double bond GO
structure model. Recently, our x-ray photoelectron spectroscopy
(XPS) simulation also led to a prediction of the epoxy pair and
epoxy-hydroxy pair species in highly oxidized GO
samples.\cite{Zhang200905}

Nuclear magnetic resonance (NMR) is the most widely used experimental
technique in GO structure study. \cite{Mermoux199169, He199654,
Lerf199757, Lerf199877, Cai200815, Gao200903} In fact, the popular
Lerf model is mainly based on $^{13}$C NMR experiments.
\cite{Lerf199877} Currently, very high quality NMR spectrum is
available via synthesis of almost fully $^{13}$C-labeled GO.
\cite{Cai200815} There are mainly three broad resonances in the
$^{13}$C NMR spectrum of GO. By checking a series of reactions of GO with
different reagents, the peak around 60 ppm is assigned to carbon atoms bonding
to the epoxy group, and the peak around 70 ppm is corresponding to
the hydroxy group connected carbon atoms. \cite{Lerf199877} The sp$^2$
carbon has a NMR peak at about 130 ppm. Besides the three main peaks,
there are also three small peaks clearly shown in the high
resolution $^{13}$C NMR spectrum. \cite{Cai200815} Gao et al. \cite{Gao200903}
suggested the these three peaks at 101, 167, and 191 ppm are
corresponding to lactol, carboxy, and ketone groups, respectively.
All these NMR assignments are based on
chemical intuition, it is highly desirable to confirm them by 
first-principles simulations.

Theoretical NMR simulation has been widely used in studies of carbon
nanotubes (CNTs). \cite{Marques200633, Zurek200695, Lai200817,
Zurek200730, Zurek200844, Zurek200917} $^{13}$C NMR isotropic
chemical shifts of pristine CNTs, based on the
density functional theory (DFT) and an infinite CNT model,
depend on the electronic structure and the diameter but not on the
chirality. \cite{Lai200817} It has also been demonstrated that a wealth of knowledge
on functionalizations \cite{Zurek200730, Zurek200917} and defects
\cite{Zurek200844} of CNTs can be obtained from NMR simulations. Similarly,
new insights about GO structure are
expected to be obtain by a systematic computational
NMR study.

\section{Computational Details}

All of our calculations were performed with a plane wave based DFT
implementation within the CASTEP package, \cite{Segall200217, Clark200567}
and the gauge including projector-augmented
plane-wave (GIPAW) method \cite{Pickard200101, Yates200701} was used to
calculate NMR shielding tensors. Ultrasoft pseudopotentials
\cite{Vanderbilt199092} were used to describe the interaction
between valence electrons and ions. The revised
Perdew-Burke-Ernzerhof (RPBE) density functional \cite{Zhang199890} used
in this study has been demonstrated to be accurate enough for NMR simulation
of CNTs. \cite{Zurek200695} The energy cutoff of plane wave basis set
in our calculations was chosen to be high enough (500 eV) to converge energy,
geometry, and nuclear shielding tensor.

\begin{figure}[tbhp]
\includegraphics[width=8cm]{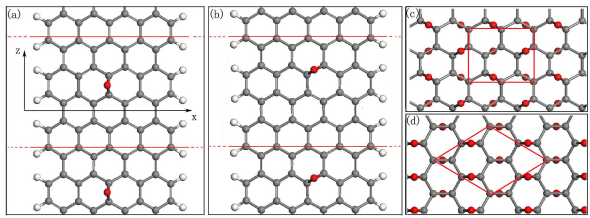}
\caption{An epoxy group adsorbs at (a) a P-site or (b) a D-site on 9-AGNR.
(c) and (d) Two full-oxidized GO with epoxide only.
Supercells used in our calculations are marked by red line.
Carbon is in gray, oxygen in
red, and hydrogen in white. } \label{fig:geo}
\end{figure}

When considering isolated oxidation groups, to avoid handling
metallic system, we adopted an armchair graphene nanoribbon (AGNR) with
a finite band gap as our substrate system.  The vacuum space between two neighboring AGNRs
was wider than 10 \AA\ in both directions perpendicular to the ribbon direction (the
$z$ direction in Figure \ref{fig:geo}a). Along $z$ direction, unless otherwise specified, a two-unit
supercell (8.52 \AA\ long according to the graphene
lattice constant) was used to make sure that oxidation groups on
the ribbon do not interact with their images in neighboring
supercells. Atom coordinates of the functionalized ribbons
were fully optimized while the supercell size was fixed. A
1$\times$1$\times$m $k$-grid was used for Brillouin zone integration,
and $m$ was large enough to make the isotropic chemical shifts of
all carbon atoms differ less than 1 ppm to those with a
1$\times$1$\times$m-1 grid.

For highly oxidized GO samples, graphene based two
dimensional (2D) models were used. And some special structure models proposed in
previous energetics studies were taken for NMR calculations. We
optimized both atom coordinates and cell parameters for all 2D
models. The $k$-grid was also carefully tested to make sure that the NMR chemical
shifts of all carbon atoms have converged.

As a commonly used method, the
calculated $^{13}$C chemical shift of benzene was used to calibrate
the chemical shift for other systems, \cite{Marques200633, Zurek200695} which generally leads to a
quantitative agreement between theory and experiment. The isotropic
chemical shift of benzene molecule calculated using a cubic cell of
15$\times$15$\times$10 \AA$^3$  was 42.2 ppm, agreeing well with previous
studies. \cite{Zurek200695} As an additional test, we calculated the chemical
shift of a (7,0) CNT, and the result, 136.3 ppm, also agrees well with the value
(136.4 ppm) obtained by Zurek et al.. \cite{Zurek200695}

\section{RESULTS AND DISCUSSION}

\subsection{Epoxy groups}

First, we consider an isolated epoxy group on an AGNR with 9 carbon chains
(9-AGNR). The epoxy group can be attached to a C-C
bond either parallel to the ribbon direction (P-site, Figure
\ref{fig:geo}a) or diagonal to the ribbon direction (D-site, Figure
\ref{fig:geo}b). C$_E$ is used to name carbon atoms
connecting to an epoxy group. The optimized C-C
bond length between the two C$_E$ carbon atoms is 1.50 \AA\ for the P-site
and 1.52 \AA\ for the D-site, close to that of $sp^3$ hybridized
carbon. For a P-site adsorbed epoxy group, chemical shifts of the
corresponding two C$_E$ atoms are both 68.0 ppm. \cite{EPAPS} Due to the its lower symmetry
and the finite AGNR width, there is a small difference (less than 1.5 ppm)
between the chemical shifts of the two C$_E$ atoms in the D-site case. Their averages
is 69.4 ppm. It is very interesting to note that
the chemical shift of isolated epoxide group is close to the experimental peak value of hydroxy group
around 70 ppm instead of 60 ppm assigned for epoxy group.

To check the width effect of the ribbon model, we have also
tested wider AGNR with 10 carbon chains (10-AGNR).
The corresponding C-C bond lengths are 1.50 and 1.53 \AA\ for the
P- and D-site, respectively.
The average chemical shifts of the two
C$_E$ atoms are 67.2 and 73.2 ppm for these two adsorption sites.
Their difference becomes larger than that of 9-AGNR, which may due to
the 3-family behavior of the ANGR electronic structure. \cite{Son0603}
However, the overall picture for 10-AGNR is the same as before,
and chemical shifts close to 70 ppm are still obtained for individual epoxy groups.

It is interesting to see how the chemical shift changes upon the change
of the epoxy group concentration. As another limit compared to the isolated
epoxide case with the lowest concentration, we consider the fully epoxidized GO, which
has the highest epoxy group concentration. Two fully-oxidized models
with chemical formula C$_2$O have been considered. Both have an optimized
C-C bond length of about 1.50 \AA. The first one (Figure \ref{fig:geo}c)
has a C$_E$ chemical shift of 59.8 ppm. The second one (Figure \ref{fig:geo}d) is part of
a previously proposed GO structure model, \cite{Yan200902} which leads to
a C$_E$ chemical shift of about 57.4 ppm. Therefore, the 100\% functionalized
graphene epoxide has a chemical shift close to the experimental value 60 ppm,
which is much lower than that of isolated epoxides. $^{13}$C chemical shift
of epoxide in GO is thus very sensitive to chemical environment, and isolated epoxy groups
should not be widely existed in the experimentally prepared GO samples.

\subsection{Hydroxy groups}

\begin{figure}[tbhp]
\includegraphics[width=8cm]{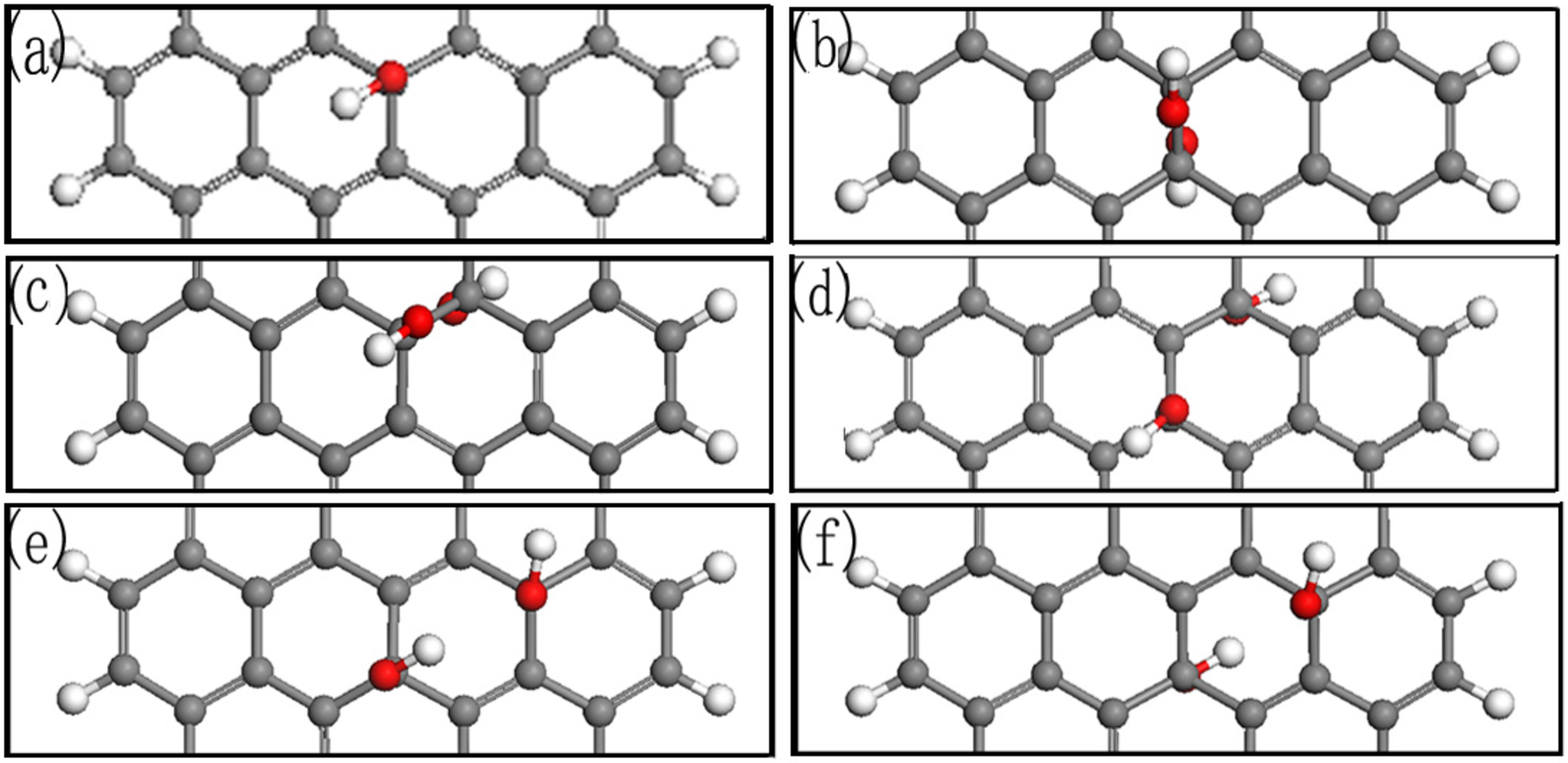}
\caption{(a) An isolated hydroxy group, (b) a P-site
hydroxyl pair, (c) a D-site hydroxyl pair, (d) a 1,3-
hydroxyl pair, (e) a 1,4-hydroxyl pair on the same side, and (f)
a 1,4-hydroxyl pair on two opposite sides. Only part of the supercell used
in calculations is shown.\cite{EPAPS} Carbon is in gray, oxygen in red and
hydrogen in white. } \label{fig:oh}
\end{figure}

For single hydroxy groups, a four unit supercell along the ribbon direction
with two hydroxy groups has been used to calculate their chemical shifts. \cite{EPAPS}  C$_H$ is used
to indicate carbon atoms connected to a hydroxy group. For 9-AGNR (Figure \ref{fig:oh}a),
the C$_H$ chemical shift of the isolated hydroxy groups is 72.4 ppm,
close to the 70 ppm peak position in experiments. 10-AGNR gives
a similar result, with a C$_H$ chemical shift of 72.2 ppm.

Isolated hydroxy pairs have also been considered. Energetically, the most stable
configuration has two hydroxy groups occupying a
1,2-site on two opposite sides of the graphene plane. P-site and
D-site can be similarly defined as in the isolated epoxy group case.
For 9-AGNR, the chemical shifts of the two C$_H$ atoms are very
close, and their averages are 72.2 and 72.0 ppm in the P-site (Figure
\ref{fig:oh}b) and the D-site (Figure \ref{fig:oh}c) cases, respectively. Both
values are close to 70 ppm assigned to hydroxy
groups in experiment. The optimized C$_H$-C$_H$ bond length is around 1.54 \AA\ for
both the P- and D-sites. Similar results are obtained for wider 10-AGNR, and the
averaged C$_H$ chemical shifts are 68.2 and 73.3 ppm for the P-site and D-site,
respectively.

To further understand the chemical shift of isolated hydroxy pair,
three other models have also been considered, where the two hydroxy
groups occupy a 1,3-site on two opposite sides (Figure
\ref{fig:oh}d), a 1,4-site on the same side (Figure
\ref{fig:oh}e), or a 1,4-site on two opposite sides (Figure
\ref{fig:oh}f). The 1,3-hydroxyl model with hydroxy groups on the
same side is not stable after optimization. For 9-AGNR, the chemical
shifts of the two C$_H$ atoms in the opposite-side 1,3-hydroxyl case
is close, and their average is 70.8 ppm. The 1,4-hydroxy groups on the
same side lead to much lower chemical shifts,  64.3 ppm for the C$_H$ atom
closer to the ribbon edge and 63.0 ppm for the other C$_H$ atom. If wider
10-AGNR is used, the corresponding chemical shifts are 67.1 and
64.5 ppm. For 1,4-hydroxy groups on two opposite
sides, the averaged chemical shifts are 63.4 and 65.4 ppm for 9-AGNR and 10-AGNR, respectively.

Therefore, isolated hydroxy group and hydroxy pair in many cases give C$_H$
chemical shifts around 70 ppm. But in some other cases (1,4-hydroxyl), it is also possible to
obtain chemical shifts  between 60-70 ppm.
At the fully-oxidized limit, hydroxyl groups will
bond to two opposite sides for each neighboring carbon atom pair.
However, there is still a large repulsion between hydroxy groups. \cite{Nu0941}
The optimized C-C bond length is 1.64 \AA, which is
much larger than a normal C-C bond length and leads to an
energetically very unfavorable structure. The calculated C$_H$ chemical
shift is 91.7 ppm.

\subsection{Mixed epoxy and hydroxy groups}

In real GO samples, both epoxy and hydroxy groups exist, and
they are expected to be randomly distributed on the graphene basal planes.
Therefore, a realistic GO model must include both epoxy and
hydroxy groups and consider the interaction between them. For simplicity, we still
start from the lowest concentration limit using ribbon models.
Obtained structure motifs can be used as building blocks for a more
realistic GO model with higher oxidation group concentration.

\begin{figure}[tbhp]
\includegraphics[width=8cm]{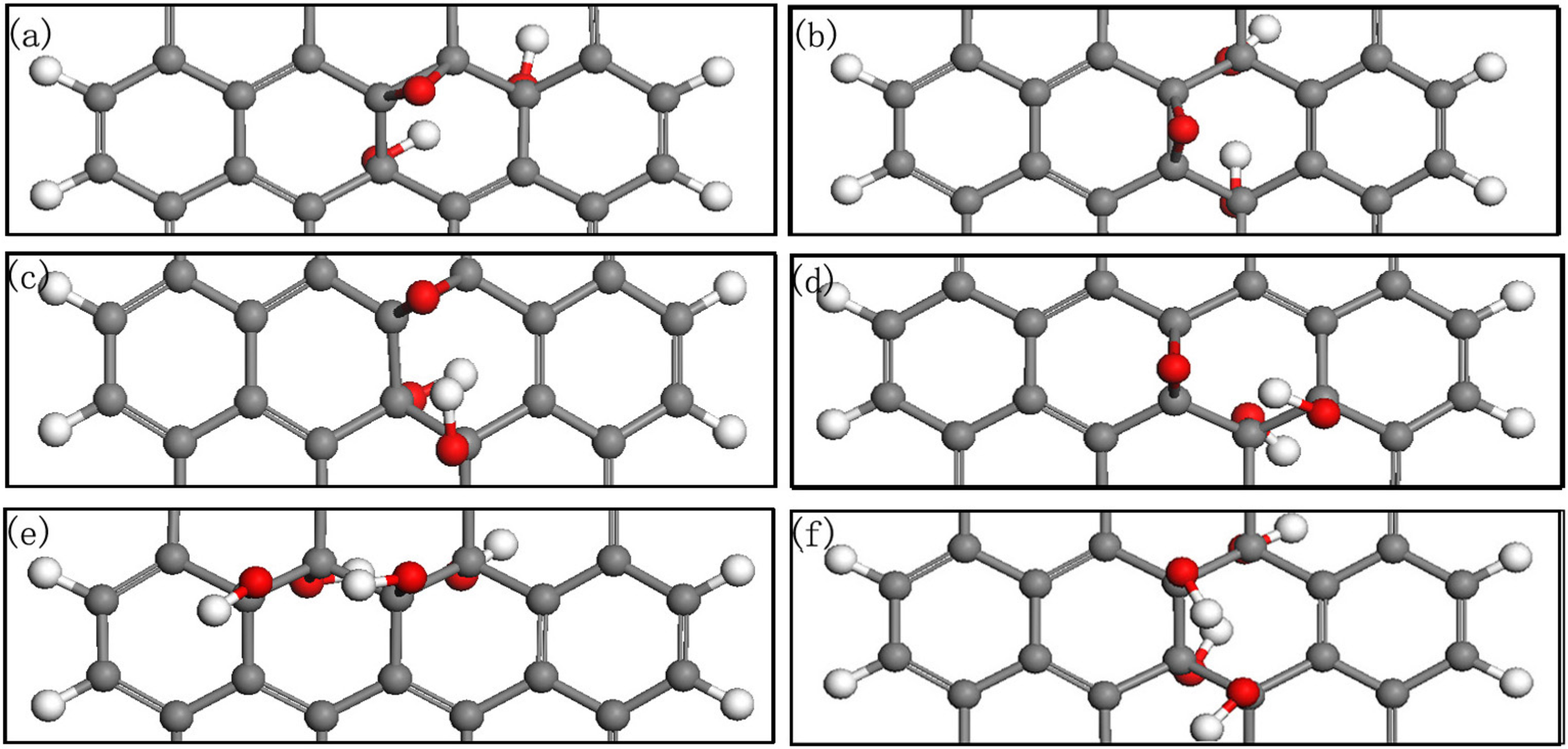}
\caption{Structure A locates at (a) a D-site or (b) a P-site in 9-AGNR.
Structure B locates at (c) a D-site or (d) a P-site in 9-AGNR.
Two hydroxyl pairs in (e) zigzag or (f) armchair directions. Only part of the supercell used
in calculations is shown. Carbon is in gray, oxygen
in red, and hydrogen in white. } \label{fig:acs}
\end{figure}

Based on energetics consideration, two stable local structures
consisting both epoxide and hydroxyl have been proposed.
\cite{Wang200995}  Here we adopt them in our ribbon model and name
them structure A and structure B (Figure \ref{fig:acs}),
respectively. When structure A is put on 9-AGNR at a D-site (Figure \ref{fig:acs}a), the
chemical shifts of two C$_E$ atoms are close, with an average of 56.4
ppm. The distance between the two epoxide carbon atoms is 1.47 \AA,
slightly shorter than those of isolated epoxide. Chemical shifts for
the two C$_H$ atoms are 66.4 and 70.5 ppm, with the one closer to ribbon
edge 4.1 ppm larger. This trend is in agreement with
isolated 1,4-hydroxyl pair. When structure A is located at a P-site (Figure
\ref{fig:acs}b), the average C$_E$ chemical shift is 58.2 ppm.
The optimized C-C bond length in the epoxide is still 1.47 \AA.
The chemical shifts for the two
C$_H$ atoms are 69.3 and 65.5 ppm, with the up C$_H$ about 4 ppm
larger. This is possibly due to the hydrogen bond formed between the two
hydroxy groups.  Similar results have been obtained for 10-AGNR.

For structure B at a D-site (Figure \ref{fig:acs}c), chemical shifts
of two C$_E$ atoms are 62.2 and 69.1 ppm. While the chemical shifts of two
C$_H$ atoms are close, with an average of 70.5 ppm. Here, the bond length between two
epoxide carbon atoms is 1.49 \AA, and it is 1.56 \AA\ between the two hydroxyl
carbon atoms. For P-site (Figure \ref{fig:acs}d), the chemical shifts of
the two C$_E$ and two C$_H$ atoms are close, and their averages are 62.8
and 69.2 ppm, respectively. The distance between two epoxide carbon atoms is 1.48 \AA.
Such a bond length slightly shorter than that of an isolated epoxide may be
an indicator of lower chemical shift. In the 10-AGNR case, we get similar result for P-site,
while for D-site, the chemical shift difference for two C$_E$ atoms becomes much smaller,
with an average chemical shift of 64.7 ppm. This result indicates that the big difference
of the two C$_E$ chemical shifts observed in 9-AGNR is not intrinsic.

Now, we have seen that, by adding proximate hydroxy groups, it is
possible to decrease the chemical shift of epoxy group from
about 70 ppm for isolated group towards the experimental value at 60 ppm. Therefore, in a
realistic GO model, epoxy groups should be close to hydroxy and other epoxy groups.
With higher oxidation group concentration, many GO
structure models have been reported in the literature. \cite{Boukhvalov200897,
Yan200902, Lahaye200935, Wang200995} We choose some typical structures
for NMR simulations.

\begin{figure}[tbhp]
\includegraphics[width=8cm]{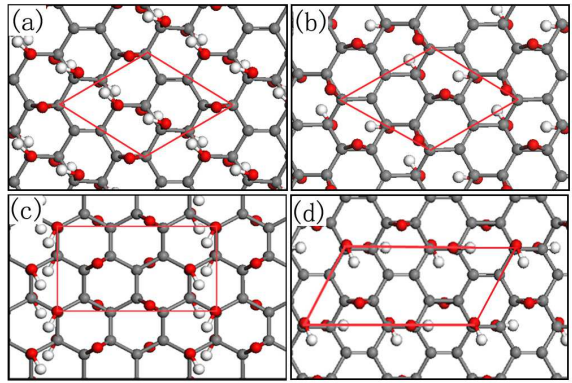}
\caption{Typical GO structure model proposed by (a)
Boukhvalov et al., (b) Lahaye et al.,
(c) Yan et al., and (d) Wang et al.. The supercells used in our calculations are
marked with red lines. Carbon is in gray, oxygen in red, and
hydrogen in white. } \label{fig:typ}
\end{figure}

Based on a 2$\times$2 supercell, Boukhvalov et al. \cite{Boukhvalov200897} proposed
the most stable structure of GO with
a 75\% coverage of oxidation groups. Their model (Figure \ref{fig:typ}a) has
a chemical formulism of C$_8$(OH)$_4$O, and its unit cell is
composed of a structure B unit and a 1,2-hydroxy
pair. An important feature of the Boukhvalov model is that the
hydroxy groups form chains, with hydrogen bonds forming within the
chains, which is a main reason why such a structure is energetically very stable.
The average chemical shifts
of C$_E$ and C$_H$ are 66.8 and 76.8 ppm, respectively. Both are
much larger than their experimental values. The C$_E$-C$_E$ bond length
is 1.51 \AA, and the averaged length of C-C bonds in
hydroxy chains is 1.58 \AA, larger than that of
isolated hydroxy pair. We find that the hydroxyl molecules are the main reason of the wrinkling of carbon
skeleton, as also suggested in a previous study. \cite{Lahaye200935} When hydroxy groups form chains, it elongates the corresponding
C-C bonds, which may then generate larger C$_H$ chemical shifts. 

With the same 2$\times$2 supercell, calculations by Lahaye et al. suggest
that too many hydroxy groups will lead to a too large tension. \cite{Lahaye200935}
Therefore, instead of C$_8$(OH)$_4$O, a GO model (Figure \ref{fig:typ}b) with a chemical composition C$_8$(OH)$_2$O$_2$ 
was suggested. In this model, hydroxyl molecules are attached to carbon atoms directly adjacent to
epoxides, but at an opposite side of the carbon plane. Hydroxy groups do not
form chains any more. Actually, this structure can be considered as a composition of
a structure A unit and an additional epoxy group per unit cell. After optimization, the length of the bond between two
epoxide carbon atoms is 1.49 \AA, and the average length of the three C-C bonds connecting to a hydroxy
group is about 1.54 \AA.  Our NMR calculations
give an average C$_E$ chemical shift of 60.6 ppm, and the chemical shifts
of the two C$_H$ carbon atoms are both 75.6 ppm. Compared to experiments, the 1,4-hydroxyl
configuration adopted in this model leads to a too large chemical shift.

Based on a systematic energetics analysis, Yan et al. argued that
all structures with a negative formation energy are fully oxidized GO
with 1,2-hydroxyl pairs forming a chain-like structure and epoxy groups
on remaining C atoms. \cite{Yan200902}
One example is C$_6$(OH)$_2$O$_2$ (Figure \ref{fig:typ}c),
where 1,2-hydroxyl chains
are along the armchair direction instead of
the zigzag direction as in the Boukhvalov model. The distance between two
C$_E$ atoms is about 1.49 \AA, and the average C$_H$-C$_H$ bond length
is 1.58 \AA, similar to those in chains in the zigzag direction.
The average C$_E$ chemical shift is 61.9 ppm, close to the experimental value.
While the average C$_H$ chemical shift is still too large (78.5 ppm), since
hydroxy groups form chains.

Another GO structure model is constructed based on the structure A building block. \cite{Wang200995}
As shown in Figure \ref{fig:typ}d, in this model, structure A units are
connected along the armchair direction. Here,
the ratio of C(sp2)/C(-O-)/C(-OH) is 1:1:1. The averaged C-C distance for epoxy groups
is 1.49 \AA, while that for hydroxy pairs is 1.57 \AA. The average chemical 
shifts for C$_E$ and C$_H$ atoms are 61.1 and 73.8 ppm, respectively.  This structure model
gives the best agreement with NMR experiments among the several models we studied here. 
We note several important structure features of this structure:
the proximity between hydroxyl and epoxide, 1,2-hydroxyl pairs but without forming a chain structure,
and a balanced ratio between hydroxyl and epoxide groups. 

\subsection{$sp^2$ carbon}

The chemical shift of $sp^2$ carbon is also very important, since
it is an indicator of the distribution of this carbon species. The chemical shifts of $sp^2$ carbon
in nanoribbon models, including clean AGNRs, are generally between
119-130 ppm, with most of them around 120-125 ppm. They are
smaller than the experiment value (129-133 ppm), which may be due to the
limited width of the nanoribbons considered in this study and the hydrogen atoms at
the ribbon edges. For $sp^2$ carbon in an isolated C-C double bond surrounding by epoxy
and hydroxy groups, we generally get a chemical shift much higher than the experimental value. 
All those GO models in Figure \ref{fig:typ}, which is not fully oxidized, have isolated C-C double bonds.
The chemical shifts of isolated C-C double bonds are 143.4, 147.2, and 137.2 ppm for the model 
in Figure \ref{fig:typ}a, \ref{fig:typ}b, and \ref{fig:typ}d, respectively. Therefore,
isolated sp$^2$ carbon pair is not the main form of sp$^2$ carbon in GO.

Yan et al. have suggested that a stable GO
structure is composed of fully-oxidized regions like that shown in Figure \ref{fig:typ}b
and $sp^2$ carbon strips between them. \cite{Yan200902} Although their fully oxidized
GO model gives too high C$_H$ chemical shifts, our calculated results 
support their description on $sp^2$ C. For aromatic strips between fully oxidized regions,
we obtain an averaged $sp^2$ chemical shift about 134 ppm, \cite{EPAPS} which is in
good agreement with experiments. We note that the large radius limit of the $^{13}$C chemical shift
of CNTs are much lower than 130 ppm. \cite{Lai200817} Therefore, 
very large graphene area is also not expected to be existed in GO. This is
consistent with the experimental cross peak between sp$^2$ C and epoxy/hydroxy groups in 2D NMR. \cite{Cai200815} 
Finally, we reach the following picture about $sp^2$ carbon in GO: the main form of $sp^2$ carbon
is small size aromatic clusters between highly oxidized regions.

\subsection{Other oxidation groups}

The assignment of the three small peaks in GO NMR spectrum is more
difficult than that of the three strong resonances. Only recently, based on
$^{1}$H-$^{13}$C cross polarization (CP) spectrum, it
was suggested that the 101 ppm signal is resulted from
non-protonated carbons, \cite{Cai200815}  
possibly a peripheral structure of GO containing
five- and six-membered-ring lactols. \cite{Gao200903} The 167 and 190 ppm peaks
have been tentatively assigned to ketone and  ester carbonyl, respectively.
\cite{Cai200815, Gao200903}

\begin{figure}[tbhp]
\includegraphics[width=8cm]{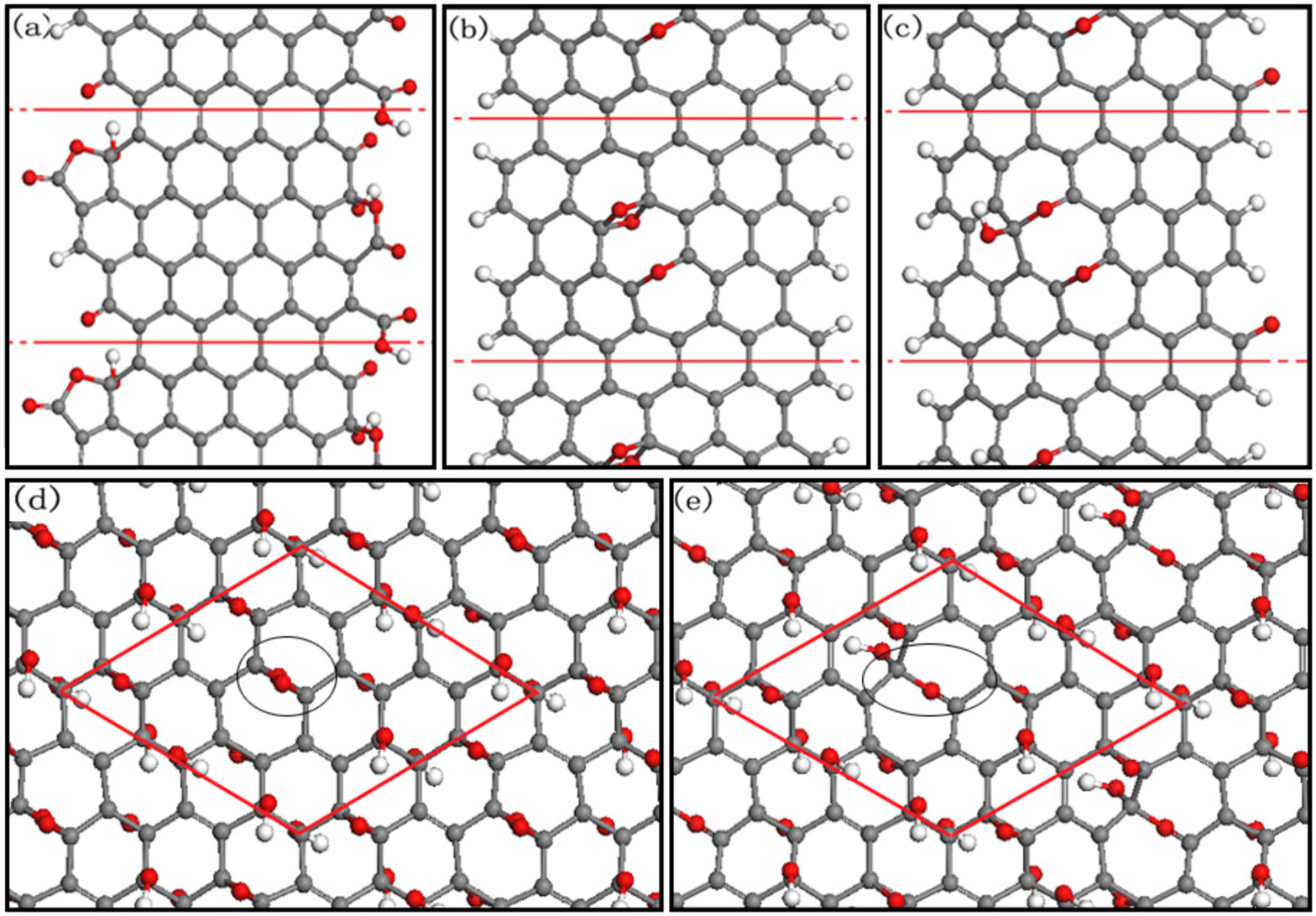}
\caption{(a) An AGNR model with 5- and
6-membered-ring lactols, carboxylic acid groups, and ketone groups. AGNR models with (b) an epoxy pair and (c)
an epoxy-hydroxy pair. Two dimensional models with (d) epoxy pair and (e)
epoxy-hydroxy pair. The unit cells used in our calculations are
marked by the red line. Carbon is in gray, oxygen in red, and
hydrogen in white. } \label{fig:self}
\end{figure}

To confirm those assignments reported in the literature, we construct an AGNR 
model (Figure \ref{fig:self}a), which contains 5- and
6-membered-ring lactols, esters, carboxylic acid, and ketone groups. 
Chemical shifts of the $sp^3$ carbon atom in
5- and 6-membered-ring lactols are 104.9 and 96.9 ppm,
respectively. Both are close to 101 ppm. The averaged chemical shift of C=O is 191.6
ppm, which is also in agreement with the
190 ppm peak in previous assignments. However, we note that the chemical shift of C=O can be strongly
affected by its environment. For example, in Fig \ref{fig:self}c,
the edge C=O group has a chemical shift only 173.1 ppm.
The chemical shift of the edge carboxyl carbon is 162.2 ppm, close to
167 ppm peak in experiments. There is also a COOH structure in  5- and 6-membered-ring lactols,
which gives chemical shifts of 166.3 and 163.2 ppm
respectively. Therefore, our calculations generally confirms previous
assignment.

However, in a previous 2D NMR experiment, \cite{Cai200815} 
cross peak has been observed for the 101 ppm peak, but not the other two minor
peaks. Therefore, the 101 ppm peak is different in
the three minor peaks. It may comes from on-plane groups instead of
edge group. In a previous study, we have suggested that epoxy pair
and epoxy-hydroxy pair may exist in highly oxidized GO samples.
\cite{Zhang200905} It is interesting to see what is their NMR
signals. We use both ribbon model and 2D model to simulate
their chemical shifts. Figure \ref{fig:self}b shows an AGNR with an
epoxy pair, where both carbon atoms bound by the epoxy pair have a chemical shift around
103.7 ppm. The chemical shift for
epoxy-hydroxy pair in the ribbon model (Figure \ref{fig:self}c) is 105.9 ppm. 
Bulk models give similar results, with a 101.4 ppm average chemical shift
for epoxy pair (Figure \ref{fig:self}d) and 106.8 ppm for epoxy-hydroxy pair (Figure \ref{fig:self}e). 
Therefore, the NMR signal at 101 ppm strongly support the
existence of epoxy pair we proposed in a previous cutting mechanism
study \cite{Li200920} and XPS simulation. \cite{Zhang200905}
Although its chemical shift is also close to 101 ppm, our results can not
be used to support the existence of the epoxy-hydroxy pair, since the 101 ppm
signal disappears in CP spectrum. \cite{Gao200903}

\subsection{Discussion}

Isolated epoxide presents a C$_E$
chemical shift around 70 ppm.  It decreases to around 60 ppm in
most cases, when epoxide is in close proximity with hydroxyl. A previous study
\cite{Lahaye200935} suggested that the sole presence of 1,2-ether
epoxy group is not stable due to the accordingly created big tension on
the carbon grid, and hydroxyl is needed to safe guard the stability
of the structure. Therefore, the proximity of epoxy and hydroxy groups 
has both significant geometric effects and electronic effects. It's interesting to see whether
the change of the chemical shift of the epoxy group is mainly a geometric or an electronic
effect.

For this purpose, a computer experiment has been performed by
calculating chemical shifts of an artificial system, which has the
hydroxy groups taken off from a previously
optimized geometry but fixing all other atoms. For the structure shown in
Figure \ref{fig:acs}a, the average C$_E$ chemical
shift becomes 76.9 ppm after hydroxyl groups being taken
off. For the structure in Figure
\ref{fig:acs}b, the averaged C$_E$ chemical shift also becomes 68.8 ppm,
with obvious increase. Therefore, electronic effect plays
an important role in the hydroxy induced chemical shift decrease of
epoxide.

Our results suggest that hydroxy chain is not
widely existed in GO, since its chemical shift is too high. To
confirm this conclusion, we further consider the effect of
nearby epoxy groups. Similar calculations have been performed by
removing epoxy groups from the optimized geometries with hydroxy
chains. For the structure shown in Figure \ref{fig:typ}a, the average C$_H$ chemical
shift is 78.2 ppm after removal of epoxide, which is
only 1.4 ppm larger than the original value. For
the structure in Figure \ref{fig:typ}d, we also only get a 1.7 ppm chemical shift difference
by removing the epoxy groups.

Therefore, the great geometrical distortions brought by the hydroxy
chains should be the main reason of the chemical shift increase related to
isolate hydroxy pairs. To see if there is any other effect, we consider
two neighboring hydroxy pair. They can arrange along the zigzag
direction (Figure \ref{fig:acs}e) or along the armchair direction
(Figure \ref{fig:acs}f). In both cases, the
C$_H$-C$_H$ bond length is about 1.55 \AA, much smaller than those
in hydroxy chains. Therefore, the geometrical effect is small in
these two cases.

In the zigzag direction case, the chemical shifts are 74.5, 67.2,
72.9, and 71.6 ppm for C$_H$ carbon atoms from left to right, respectively. 
In the armchair case, the corresponding chemical shift is 74.1, 70.1, 70.4, and 74.2 ppm, 
from up to down. We can clearly see that the hydroxy group with a hydrogen 
bond pointed to it has a higher C$_H$ chemical shift. Therefore, hydrogen bonding may
also be an important factor in increasing the chemical shift of hydroxy
group in hydroxy-pair chain.

\section{conclusions}

Our results are consistent with the Lerf model of GO, where the main
species in GO are epoxide, hydroxyl, and $sp^2$ carbon. Their chemical 
shifts are very sensitive to their chemical environment, 
which leads to the result that peaks in GO NMR spectrum are very broad. 
When it is isolated, the chemical shift of epoxide is too high. Therefore, epoxide
prefers to in close proximity with hydroxyl. Chemical shift of hydroxyl groups
is closely related to its geometrical structure. Isolated 1,2-hydroxy pairs have
stable chemical shifts around 70 ppm. However, when they form hydroxy chains, the chemical
shift becomes much higher mainly due to the elongated C-C bonds. As a result, hydroxy chains
widely proposed in literature based on energetics considerations
should not be an important structure motif of GO. Our calculations also confirmed
the existence of aromatic carbon clusters among highly oxidized regions.
Considering its cross peak in 2D NMR, the small peak at 101 ppm is more likely contributed
by on-plane groups, such as epoxy pair we proposed earlier.

\end{document}